\documentstyle[aps]{revtex}

\newcommand\X{{\mathrm{X}}}

\newcommand\T{{\mathrm{T}}}
\newcommand\x{x}
\newcommand\y{y}
\begin{document}
\title{\Large\bf Spacetime states and covariant quantum
theory} \author{Michael Reisenberger ${}^{ab}$ and Carlo
Rovelli ${}^{ac}$\\[.1cm] {\it ${}^{a}\!\!$ Centre de Physique
Th\'eorique, Luminy, F-13288 Marseille, EU}\\
{\it ${}^{b}\!\!$ Physics Department, Pittsburgh
University, PA-15260, USA}\\
{\it${}^{c}\!\!$ Facultad de Ciencias, Universidad de la Rep\'ublica,
Igu\'a 4225, Uruguay}}
   \date{\today}  \maketitle
 \begin{abstract}

In it's usual presentation, classical mechanics appears to give time a
very special role.  But it is well known that mechanics can be
formulated so as to treat the time variable on the same footing as the
other variables in the extended configuration space.  Such covariant
formulations are natural for relativistic gravitational systems, where
general covariance conflicts with the notion of a preferred
physical-time variable.  The standard presentation of quantum
mechanics, in turns, gives again time a very special role, raising
well known difficulties for quantum gravity.  Is there a covariant
form of (canonical) quantum mechanics?

We observe that the preferred role of time in quantum theory is the
consequence of an idealization: that measurements are instantaneous. 
Canonical quantum theory can be given a covariant form by dropping
this idealization.  States prepared by non-instantaneous measurements
are described by ``spacetime smeared states".  The theory can be
formulated in terms of these states, without making any reference to a
special time variable.  The quantum dynamics is expressed in terms of
the propagator, an object covariantly defined on the extended
configuration space.

  \end{abstract} \vskip1cm

\section{Introduction}

In this paper, we discuss a covariant formulation of canonical quantum
mechanics.  This formulation is based on the propagator and on a
representation of quantum states which we will call
``spacetime-smeared quantum states".  We think that this formalism can
play a role in several problems, such as for instance: the
interpretation of spinfoam quantum gravity, the interpretation of
quantum cosmological models, certain interpretational issues in the
quantum mechanics of a single relativistic particle and the problem of
the computation of the time of arrival in quantum mechanics.  These
problems are all connected, and refer to the role that time plays in
the formalism of quantum theory.  Ideas closely related to the ones
presented here can be found in the work of Jim Hartle
\cite{Hartle,hartle2} and Don Marolf \cite{Marolf,detector}.  But also
in Bryce DeWitt \cite{DeWitt}, Chris Isham \cite{isham}, John Klauder
\cite{Klauder}, Jonathan Halliwell \cite{JH}, Rodolfo Gambini and
Rafael Porto \cite{Gambini} and certainly others. 

Our main tool is the spacetime-smeared representation of quantum
states, which we define in Section \ref{free}.  This representation is
the natural one for states prepared in measurements which are not
instantaneous \cite{hartle2} and it is spacetime symmetric.  We
discuss in detail an example of such a measurement.  We then construct
a general formulation of quantum theory based on these states, in
Section \ref{general}.  In Section \ref{applications}, we apply this
formulation to several concrete situations in order to test its
viability and to point out its advantages.  This formulation clarifies
some issues in the formulation of the quantum theory of a single
relativistic particle, and in relation to the time of arrival problem,
and it helps us to give a consistent interpretation to quantum
cosmological models.  In Section \ref{issues}, we discuss various
conceptual issues raised by this formulation and in Section
\ref{concl} we briefly summarize.

One of the motivations for the present work is to help define the
interpretation --that is, the relation between formalism and
observation-- of the formalisms for quantum gravity developed by the
authors and others (see \cite{spinfoam} and references therein) in
which the central object that is computed is the propagator (or
``projector on physical states") $P$.  This work develops ideas on
generally covariant quantum theory previously published by one of us
(CR); however, the tools presented here allow a considerable
simplification.

Finally, a note on notation: we use capital roman letters $(\X,\T)$
for the space and time coordinates, while we use lower case italic
letters $\x,\y,\ldots$ for spacetime points, and later on, for points
in the extended configuration space.  Thus for a particle in two
dimensions ${\x }=({\X},{\T})$.

\section{Spacetime smeared quantum states}
\label{free}

Consider a free, non-relativistic particle in one space dimension. 
Let $\psi({\X},\T)$ be its Schr\"odinger wave function, namely a
solution of the free Schr\"odinger equation
\begin{equation}
    \imath\hbar {\partial\over\partial \T}\ \psi(\X,\T)
    =-{\hbar^{2}\over 2m}\ {\partial^{2}\over\partial {\X}^{2}}\
    \psi(\X,\T).
    \label{eq:Sch}
\end{equation}
The Hilbert space of the quantum theory is the space of normalizable
solutions to the Schr\"odinger equation.  It can be represented by the
space $L_{2}[R]$ of square integrable functions on space alone.  The
wavefunction $\psi(X,T)$ is represented by the square integrable
function $\Psi({\X})=\psi({\X},0)$ at a fixed time ${\T}=0$, and we
will often denote the state by $|\Psi\rangle$.  In this representation
the scalar product is
\begin{equation} 
    \langle\Psi|\Psi'\rangle=\int d\X\ \overline{\Psi(\X)}\Psi'(\X).
\end{equation}

The spacetime wavefunction $\psi$ can be reconstructed from $\Psi$
using the propagator.  We denote the (generalized) eigenstates of the
position operator ${\X}$ by $|{\X}\rangle$ and the generalized
eigenstates of the unitarily evolving Heisenberg position operator
$X({\T})$ as $|{\X},{{\T}}\rangle$ (so that
$|{\X}\rangle=|{\X},0\rangle$)
.  Thus $\Psi({\X})=\langle {\X}
|\Psi\rangle$ and $\psi({\X},{\T})=\langle {\X}, {\T} |\Psi\rangle$. 
The propagator of the Schr\"odinger equation is
\begin{eqnarray} 
W({\X},T;{\X}',{\T}') &=&\langle {\X},{\T}|{\X}',{\T}'\rangle \nonumber \\
&=& \int  \frac{dp}{2\pi\hbar}\ dE 
\ e^{i/\hbar[p({\X}-{\X}')-E({\T}-{\T}')]}\ \delta(E-p^{2}/2m)\nonumber \\
&=& \int \frac{dp}{2\pi\hbar} 
\ e^{i/\hbar[p({\X}-{\X}')-p^{2}/2m({\T}-{\T}')]} \nonumber \\
&=& \left(2\pi m\over i\hbar({\T}-{\T}')\right)^{1\over 2}\
\exp\left\{-{m({\X}-{\X}')^{2}\over 2i\hbar({\T}-{\T}')}\right\}.
\end{eqnarray}
When viewed as a function of $X$ and $T$, with $X'$ and $T'$ held
fixed, this is a solution of the Schr\"odinger equation which at time
$\T={\T}'$ is given by a delta distribution in $\X-{\X}'$.  Each
function $\Psi({\X})$ determines a solution of the Schr\"odinger
equation by
\begin{equation} 
    \psi({\X},{\T}) = \int d{\X}'\ W({\X},{\T};{\X}',0)\  \Psi({\X}').
\end{equation} 
Thus the wavefunctions allowed by the Schr{\"o}dinger equation can be
characterized by the functions $\Psi(X)$ of space only.

Now we shall consider another representation of quantum states.
Consider the wavefunction given by 
\begin{equation} 
  \psi_{f}({\X},{\T}) \equiv \int d{\X}' d{\T}'\ 
  W({\X},{\T};{\X}',{\T}')\ 
  f({\X}',{\T}').
   \label{mapP}
\end{equation} 
where $f({\X},{\T})$ is a smooth function on spacetime.  The wave
function $\psi_{f}({\X},{\T})$ is a solution of the Schr\"odinger
equation as well.  In the standard ``instantaneous" representation
discussed above, the wavefunction is represented by the function of
space obtained by restricting $\psi_{f}({\X},{\T})$ to ${\T}=0$.  If
the restriction is square integrable the wavefunction is a
normalizable state.  We use the notation $|f\rangle$ for this state. 
That is
\begin{equation}
    |f\rangle = \int d{\X} d{\T}\ f({\X},{\T})\
    |{\X},{\T}\rangle.
\end{equation}

Since the propagator satisfies the properties 
\begin{eqnarray}
    W({\X},{\T};{\X}',{\T}')^* = W({\X}',{\T}';{\X},{\T})
\end{eqnarray}
and
\begin{eqnarray}
   W({\X},{\T};{\X}',{\T}') = \int d{\X}'{}'\ 
    W({\X},{\T};{\X}'{}',{\T}'{}') W({\X}'{}',{\T}'{}';{\X}',{\T}'),
\end{eqnarray}
the scalar product between two such states is 
\begin{eqnarray}
   \langle f|f'\rangle =\int d{\X}\ d{\T} \int
    d{\X}'\ d{\T}'\ \overline{f({\X},{\T})}\
    W({\X},{\T};{\X}',{\T}')\ f'({\X}',{\T}').
    \label{bform1}
\end{eqnarray}
It is easy to see that the spacetime smeared states $|f\rangle$ are
dense in the Hilbert space of the theory.

We shall call the normalizable state $|f\rangle$ labeled by the
spacetime function $f({\X},{\T})$ a ``spacetime-smeared quantum
state'', or simply a ``spacetime state''.  It is important to notice
that this denomination refers simply to the representation that the
state is given.  It is a perfectly ordinary normalizable quantum state
in the ordinary Hilbert state of the theory.

It is clear that the spacetime-smeared representation is highly
redundant -- many different spacetime functions $f$ give rise to the
same state -- while the instantaneous representation is unique.  This
situation is a bit remeniscent of that found in electromagnetism,
where the gauge freedom can be essentially eliminated in the Coulomb
gauge, but at the cost of making the formalism non-covariant.

A particular class of spacetime states plays an important role in what
follows.  These are the spacetime states associated to small spacetime
regions $\cal R$.  Here ``small" means smaller than any spatial or
temporal scale involved in the problem being studied.  We define the
state $|{\cal R}\rangle$ associated to a spacetime region $\cal R$ as
the normalized spacetime state defined by the characteristic function
of ${\cal R}$, that is
\begin{equation}
    |{\cal R}\rangle = C_{\cal R} \int_{\cal R} d{\X}d{\T}\ \
    |{\X},{\T}\rangle.
    \label{Rstates}
\end{equation}
The normalization factor is easily computed as 
\begin{equation}
    C_{\cal R} = \left(\int_{\cal R} d{\X}d{\T} \int_{\cal R}
    d{\X}'d{\T}'\ W({\X},{\T};{\X}',{\T}')\right)^{-{1\over 2}}. 
    \label{Rnorm}
\end{equation}

In the rest of this section we study the physical interpretation of
the states $|f\rangle$ and the states $|{\cal R}\rangle$.  In the rest
of the paper we will use this spacetime-smeared representation and the
propagator to construct a covariant formulation of canonical quantum
theory.

Our central claim is (i) that the spacetime-smeared states $|f\rangle$
are natural objects, once one drops the unrealistic idealization that
measurements are instantaneous, and (ii) that they make a spacetime
symmetric formulation of quantum theory possible.

\subsection{Real measurements}\label{realistic}

Roughly speaking, if we measure the position of the particle at time
$\T=0$, and we find the particle in ${\X}=0$, we can then assume that
the particle is in the state $|{\X}\rangle$.  However, as is well
known, no real measuring device can resolve a particle's position with
infinite precision.  Every real measuring device has a finite
resolution $a$.  We can represent a particle that at $\T=0$ has been
detected in ${\X}=0$ by an apparatus with spatial resolution $a$ by a
wave packet spread over a finite region of size $a$.  Its state will
have the form
\begin{equation}
    |\Psi\rangle = \int d{\X}\ f({\X})\  |{\X}\rangle, 
\end{equation}
where $f({\X})$ is, say, a function with support in the interval 
$[{\X}-a,{\X}+a]$, or, perhaps, a Gaussian smearing function 
\begin{equation}
    f({\X}) = e^{-{{\X}^{2}\over 2a}}. 
\end{equation}

What about time?  In the usual textbook description, the measurement
is assumed to be instantaneous.  We observe, instead, that a real
measuring device interacts with the system being measured for a finite
interval of time.  A real measurement never refers to a single sharply
defined time \cite{hartle2}.  Thus, in the case of a position
measurement neither the position nor the time at which the particle is
seen is resolved with infinite precision.  Let's say that the
measuring device resolves the time with precision $\epsilon$.  We
would like to claim that a particle detected in ${\X}=0$ at ${\T}=0$
with an experimental device having space resolution of order $a$ and
time resolution of order $\epsilon$ can be described by a wave packet
with the form 
\begin{equation}
    |f\rangle = \int d{\X} d{\T}\ f({\X},{\T})\  |{\X},{\T}\rangle, 
    \label{state}
\end{equation}
where $f({\X},{\T})$ is a function concentrated in the region
$[-a,+a]\times [-\epsilon,+\epsilon]$, such as for instance, the
characteristic function of the region; or, say, by a Gaussian smearing
function
\begin{equation}
    f({\X},{\T}) = e^{-{{\X}^{2}\over 2a}-{{\T}^{2}\over 2\epsilon}}. 
\end{equation}
Notice that in equation (\ref{state}), the state is naturally
represented as a spacetime-smeared state, as defined in the previous
section.

In order to clarify this point, we now describe a simple model of a
measurement procedure.  For related constructions, see also
\cite{detector}.  This measurement procedure we describe is realistic
in the sense that the physical interaction responsible for the
measurement is not idealized away.  We want to measure the position of
the particle at a certain time.  That is, we want to check whether the
particle is present at a certain point $\X=0$ at a certain time
$\T=0$.  We thus set up a physical apparatus that interacts with the
particle.  This apparatus will have a pointer that tells us whether or
not the particle has been detected.  We now exploit the freedom in
choosing the boundary between the quantum system under observation and
the measuring apparatus: we treat the particle and particle detector
as the system, and consider that the Copenhagen ``measurement" is
realized when the position of the pointer is observed.  (In the
measurement of the pointer, the time duration of the measurement is
not an issue, because the pointer is static after the interaction with
particle is complete.)  This trick allows us to understand which
precise aspect of the particle state is probed by an apparatus
measuring the localization of the particle.

Let us consider for simplicity a pointer which has two possible
states.  A state $|0\rangle$, which corresponds to no detection, and a
state $|1\rangle$, which corresponds to detection.  We then represent
the state space of the coupled particle-detector system by the Hilbert
space $H_{PD}=H\otimes C^2$, where $H$ is the Hilbert space of the
particle and $ C^2$ is the state space of a two-state system.  We
write a state of the combined system as
\begin{equation}
    \Psi_0({\X})\otimes |0\rangle +
    \Psi_1({\X})\otimes |1\rangle .  
\end{equation}
The free hamiltonian of the particle is $P^2/2m$, and we take the free
hamiltonian of the detector to be zero.  We need an interaction
hamiltonian $H_{int}$, representing the interaction that gives rise to
the measurement.  $H_{int}$ must have the following properties. 
First, it must cause the transition $|0\rangle \rightarrow |1\rangle$. 
Second, the particle should interact only at or around the spacetime
position ${\X}=0, \T=0$.  Thus the interaction hamiltonian must be
time dependent, and vanish for late and early times.  We have to
concentrate the interaction around $\T=0$.  However, we cannot have a
perfectly instantaneous interaction because this would require
infinite force.  We must therefore assume that the interaction is non
vanishing for a finite period of time.  Putting these requirements
together, and requiring also that the hamiltonian is self-adjoint, we
arrive at an interaction hamiltonian of the form
\begin{equation}
 H_{int} = \alpha\ V({\X},{\T})\  \Big(\ |1\rangle \langle 0 | +|0\rangle
 \langle 1 | \ \Big)
\label{inteham}
\end{equation}
where $\alpha\ V({\X},{\T})$ is the potential acting on the particle
in the interaction (with $\alpha$ a coupling constant).  The potential
$V({\X},{\T})$ is concentrated in a finite spacetime region $\cal R$,
which we take to be concentrated around ${\X}=0$ and ${\T}=0$.  

The Schr\"odinger equation for the spacetime wavefunctions of the 
particle states $\Psi_{0}$ and $\Psi_{1}$ reads 
\begin{eqnarray}
    \imath\hbar {\partial\over\partial {\T}}\ \psi_{0}({\X},{\T})
    &=&-{\hbar^{2}\over 2m}\ {\partial^{2}\over\partial
    {\X}^{2}}\psi_{0}({\X},{\T}) +\alpha V({\X},{\T})\psi_{1}({\X},{\T})  \\
    \imath\hbar {\partial\over\partial {\T}}\ \psi_{1}({\X},{\T})
    &=&-{\hbar^{2}\over 2m}\ {\partial^{2}\over\partial
    {\X}^{2}}\psi_{1}({\X},{\T}) + \alpha V({\X},{\T})\psi_{0}({\X},{\T}).
\end{eqnarray} 

Now assume that at some early time ${\T}_{in}<<0$ the particle is in some
initial state $\psi({\X},{\T}_{in})$ and the pointer is in the state
$|0\rangle$.  What is the state of the system at a later time
${\T}_{fin}>>0$?  It is straightforward to integrate the evolution
equations to first order in $\alpha$. One obtains 
\begin{eqnarray}
     \psi_{0}({\X},{\T})\!\! & = &\! \int d{\X}'\
     W({\X},{\T};{\X}',{\T}_{in}) \ \psi_{0}({\X}',{\T}_{in}) . 
     \\
     \psi_{1}({\X},{\T}_{fin})\!\!  & = & \!  \frac{\alpha}{i\hbar}
     \int_{\cal R} d{\X}'d{\T}'\ W({\X},{\T}_{fin};{\X}',{\T}') \
     V({\X}',{\T}')\ \psi_{0}({\X}',{\T}').
    \label{evolut}
\end{eqnarray}
If the pointer is observed in the state $|1\rangle$ after the
interaction, the state of the system collapses to
$\Psi_{1}\otimes|1\rangle$.  After the measurement, the state of the
particle is thus described by the wave function (\ref{evolut}).  But
notice that this has precisely the form (\ref{state}) of a spacetime
smeared state, where the spacetime smearing function has support in
$\cal R$: 
\begin{equation}
    f({\X},{\T}) = V({\X},{\T})\ \psi_{0}({\X},{\T}). 
\end{equation}
The result to all orders in perturbation theory is more complicated,
but it is easy to see that $\psi_1$ still has the form of a spacetime
smeared state with $f$ supported in $\cal R$.  Thus we can conclude
that if we prepare a state concentrated around ${\X}=0, \T=0$ by means
of a physical measurement procedure like the one described, we
necessarily obtain a spacetime-smeared state defined by a function
$f({\X},{\T})$ with support in a finite region $\cal R$ around
${\X}=0, {\T}=0$.\footnote{The apparatus we have described is
effective for {\em preparing\/} a quantum state concentrated around
the origin, but performs poorly as a {\em detector\/} of whether a
particle is or not around the origin.  This is because there is always
a finite probability for the pointer to ``forget", namely to jump back
from $|1\rangle$ to $|0\rangle$, after the first detection.  This does
not happen to first order in $\alpha$, but it is easy to see that it
happens to second order.  We can minimize the probability of
``forgetting" by taking $\alpha$ small, but this gives us a detector
which works correctly, but has low efficiency.  Real detectors,
however, are dissipative.  For instance, in the silver nitrate
crystals of photographic film, the microscopic signal that is being
detected triggers the fall of the detector to a lower energy, with the
energy liberated being absorbed by the environment, raising it's
entropy.  There is one, or a small number, of states corresponding to
no detection, but a large number of states corresponding to detection. 
It follows that once the detector (and environment) has interacted
with the microsystem it cannot find it's way back to it's initial
state in a second interaction, even though this is energetically
possible, for reasons of statistics.  A simple way of taking this fact
into account in the model is to replace (\ref{inteham}) with the
interaction hamiltonian
\begin{equation}
 H_{int} = \alpha\ V({\X},{\T})\  |1\rangle \langle 0 | 
\end{equation}
which is not self-adjoint.  With this hamiltonian, the solution
(\ref{evolut}) is indeed exact not just to first order, but to all
orders in $\alpha$: once the detector has detected the particle, it
does not forget it.  We are thus not forced to take $\alpha$ small,
and we can use such a detector not only to {\em prepare\/} a quantum
state concentrated around the origin, but also to efficiently check
whether a particle is at the orgin or not.}%
The size of $\cal R$ is determined by the accuracy of the measuring
apparatus in resolving distances {\em and time intervals}.

Consider now an ideal case in which the region $\cal R$ is much
smaller than the size over which the wave function $\psi(\X,\T)$
varies and the potential $V(\X,\T)$ is constant over $\cal R$.  In
this case, if the pointer is observed in the state $|1\rangle$ after
the interaction, the state of the particle collapses precisely to the
state $|{\cal R}\rangle$, defined in (\ref{Rstates}-\ref{Rnorm}).  The
dependence of the final state on the initial wave function, on
$\alpha$ and on the potential, is completely cancelled by the
normalization of the state.

The model of detector that we have described is only an example, but
we think that the conclusion that no position detector can have
infinite time resolution is true in general.  Textbook model detectors
have elements of idealization that hide the finite time resolution. 
We shall not attempt a general analysis.  We close this subsection,
however, by showing that the measurement time is finite also for the
quintessential position detector: a hole in a wall.  Consider a
physical particle in three spacial dimensions $X,Y$ and $Z$.  Suppose
we want to prepare a state in which the coordinates $Y$ and $Z$ are
concentrated in a finite (two-dimensional) interval $I$.  Then we can
take a wall in the $(Y,Z)$ plane, having a hole of dimension $I$, and
have the particle go through.  If the particle passes accross the
hole, then its $Y$ and $Z$ coordinates are in $I$.  This is the
simplest position measurement discussed in textbooks.  At which time
such a measurement happens?  In the usual textbook discussion, one
takes a state described by a plane wave moving in the $X$ direction,
namely normal to the wall.  This can be interpreted as describing a
steady flux of particles, and any time consideration is thus avoided. 
But what if we have a single particle?  Then the time of the
measurement is clearly the time $T$ at which the particle reaches the
wall and crosses (or fails to cross) the hole.  But the true state of
the particle cannot be infinitely concentrated in the $X$ direction. 
The particle will be described by a wave packet that has a finite
spread $\Delta X$ in the $X$ direction and a spread $\Delta V$ in the
velocity in the $Y$ direction.  Accordingly, the wave packet will
cross the hole during a time period of the order $\epsilon \sim \Delta
X/\Delta V$.  The source of the component of the wave function
emerging on the other side of the hole is thus concentrated in the
spacetime region ${\cal R}=(T\pm\epsilon)\times I$, and not in the
region $I$ at a fixed time $T$.  Again, we conclude that the natural
description of the state emerging from the hole is in terms of a
spacetime-smeared state, not as an eigenstate of a position projection
operator at fixed time.

\subsection{The states $|{\cal R}\rangle$ and ideal spacetime
measurements}

The interpretation of Schr\"odinger quantum mechanics is based on the
postulate that $|\psi(\X,\T)|^2$ is the {\em spatial} probability
density of finding the particle at $\X$, at time $\T$.  Equivalently,
one can consider a small (smaller than any spatial scale in the
problem) spatial interval $I=[x,x+\Delta x]$ and the normalized state
$|I\rangle$ which is constant over $I$.  This state has two
properties, first, it represents a possible state prepared by a
measurement of spatial position in $I$.  Second, the probability that
an ideal detector (a detector with efficiency 1) find the particle in
$I$ is $|\langle I|\Psi\rangle|^2$.

We are searching for a spacetime version of this interpretation.  Do
the states $|{\cal R}\rangle$ associated to {\em spacetime\/} regions
have analogous properties?  In the previous subsection we have seen
that a certain measurement prepares the state $|{\cal R}\rangle$.  Can
we also say that the probability $P$ that an ideal detector detects
the particle in $\cal R$ is $|\langle {\cal R}|\Psi\rangle|^2$ ?  We
now show that the answer is yes. 

The probability of detection of the particle is given by the norm of
the state entangled with the ``yes" position of the pointer, that is
with $|1\rangle$.  It is thus
\begin{eqnarray}
    P & =&  \int d{\X}\ |\psi_{1}({\X},{\T}_{fin})|^{2}\nonumber \\ 
     & = & {\alpha\over\hbar} \int d{\X} \int_{{\cal R}} d{\X}'d{\T}'\
     \overline{W({\X},{\T}_{fin};{\X}',{\T}')
     \psi_0({\X}',{\T}')} \ \ \ 
     {\alpha\over\hbar} \int_{{\cal R}} d{\X}'{}'d{\T}'{}'\
     W({\X},{\T}_{fin};{\X}'{}',{\T}'{}')
     \psi_0({\X}'{}',{\T}'{}').
\end{eqnarray}
Integrating in $d{\X}$ and using the properties of the propagator this 
gives 
\begin{equation}
    P = {\alpha^2\over\hbar^{2}} \int_{\cal R} d^{2} {\x } \int_{\cal
    R} d^{2} {\y }\ \overline{\psi_0({\x })}\ W({\x };{\y })\ \psi_0({\y
    }),
\end{equation}
where we have switched to the spacetime notation 
${\x }=({\X},{\T})$. 

If $\cal R$ is sufficiently small that the wavefunction $\psi_0$ of
the incident particle is well approximated in $\cal R$ by its value
$\psi_0(x)$ at a point $x=(X, T)\in \cal R$ then
\begin{equation}
P =  \left(\frac{\alpha}{\hbar C_{\cal R}}\right)^2 |\psi_0(x)|^2.
\end{equation}
On the other hand the overlap of the particle's state with the
characteristic state $|\cal R\rangle$ is (always in the small region 
limit)
\begin{equation}
\langle{\cal R}|\Psi\rangle = C_{\cal R} \int_{\cal R} \psi_0(X,T)\:
dXdT = C_{\cal R} V_{\cal R} \psi_0(x),
\end{equation}
where $V_{\cal R}$ is the spacetime volume of $\cal R$.
Thus the the probability of detection is
\begin{equation}
P = \gamma |\langle {\cal R} | \Psi\rangle|^2,
\end{equation}
where
\begin{equation}
    \gamma = \left(\frac{\alpha}{\hbar V_{\cal R} C_{\cal
    R}^2}\right)^2, 
    \label{eq:kappa}
\end{equation}
which depends on the detector only, can be interpreted as the
detector's efficiency.

If the particle is in the state $|\cal R\rangle$ the detector is
triggered with efficiency $\gamma$, while if the particle is in a
state orthogonal to $|\cal R\rangle$, the detector is certainly not
triggered.

The detector realizes another characteristic of an ideal
detector of the state $|\cal R\rangle$: if it is triggered, and the
particle was already in the state $|\cal R\rangle$, the detector 
leaves the particle in the state $|\cal R\rangle$.

Since we have used first order perturbation theory, the validity of
our calculation requires that $\gamma \ll 1$, as can be seen by
calculating the second order correction.  The detector we have
analysed has therefore low efficiency: it usually misses particles
which would have triggered an ideal detector with efficiency 1. 
Nevertheless the low efficiency detector can equally be used to test
the predicted values of the probability $|\langle{\cal
R}|\Psi\rangle|^2$, by simply comparing the predicted values with the
observed detection frequencies multiplied by the calibration factor
$1/\gamma$.  We can therefore define the notion of an ideal detector,
whose efficiency is one, and whose probability dectection is
$|\langle{\cal R}|\Psi\rangle|^2$.

We thus conclude that when $\cal R$ is sufficiently small the states
$|\cal R\rangle$ are prepared and are detected by a particle detector
in the spacetime region $\cal R$.  The detection probabilities of the 
ideal detector are 
\begin{equation}
    P_{\cal R} =|\langle {\cal R}|\Psi\rangle|^2.
    \label{eq:main}
\end{equation}
Equation (\ref{eq:main}) derives from the usual probabilistic
interpretation of the wave function.  In the next section we will show
that the usual probabilistic interpretation can, in turn, be derived
from (\ref{eq:main}) .  Therefore (\ref{eq:main}) is physically
equivalent to the standard interpretation in terms of the probabities
of detection in {\em spatial\/} (equal-time) regions.  However, it
does not refer to a preferred time coordinate.  Equation
(\ref{eq:main}) is a key result of this paper: we shall take it as the
central ingredient of the interpretation of the covariant formulation
of quantum theory. 

The quantity $P_{\cal R}$ for small regions $\cal R$ provide a
probabilistic interpretation of the modulus of the wave function.  The
probabilities for large $\cal R$ or $\cal R$ consisting of two small
separated components depend on the relative {\em phase\/} of the
wavefunction at different spacetime points, thus providing the
probabilistic interpretation of the (relative) phase, as in the
standard situation.  As usual, if the $P_{\cal R}$ are measured for
all $\cal R$ on (separate instances of) the same state they
characterize the wavefunction completely (up to the overall phase).

Finally, the result that we have obtained can be expressed in the
stardard operator language as follows.  A (efficiency one) measurement
of whether the particle is in the small spacetime region $\cal R$ is
represented by the self-adjoint operator
\begin{equation}
    \Pi_{\cal R} =|{\cal R}><{\cal R}|. 
    \label{eq:AR}
\end{equation}
The corresponding classical observable has value 1 on all the
solutions of the equation of motion that cross ${\cal R}$, and zero
elsewhere.

\subsection{Relation with the spatial probability density}

We have seen above that starting from standard quantum theory one can
derive the interpretation of $P_{\cal R} = |\langle {\cal
R}|\Psi\rangle|^2$ as probability that the particle is detected in
$\cal R$.  Here, for completeness, we show how one go back from the
probability in spacetime $P_{\cal R}$ to the standard interpretation
of $|\psi|^2$ as a probability in space, and we discuss why spacetime 
probability densities cannot be defined. 

The probability to detect the particle in a small region $\cal R$ is
given by (\ref{eq:main}).  Consider small rectangular spacetime
regions $\cal R$ of spatial width $\Delta X$ and duration $\Delta T$
which satisfy the inequality $\Delta T \ll m\Delta X^2/\hbar$.  This
region can be thought as a rectangle with the timelike side much
smaller than the spacial side.  It is not hard to prove that for such
a region, and up to higher order terms in the size of the region, 
\begin{eqnarray}
     C_{\cal R}^{-2} =  \int_{\cal R} d^2x \int_{\cal R} d^2y\ 
     W(x;y) = \Delta X\ \Delta T^2
\end{eqnarray}
(because $\int_{-\infty}^{\infty} dX\ W(X,T;X',T') = 1$ and $\cal R$
is suficiently spatially wide compared to it's duration that $X$
integrals give almost the same result as the integral over
$(-\infty,\infty)$).  Consequently, for these regions we have
\begin{eqnarray}
P_{\cal R} = |\langle {\cal R}|\Psi\rangle|^2  =  C_{\cal R}^2 
\left|\int_{\cal R} \psi(x) d^2x \right|^2 
= |\psi(x)|^2\  \frac{\Delta X^2 \Delta T^2}{\Delta X \Delta T^2}	 =
|\psi(x)|^2 \Delta X
\end{eqnarray}
(where $x$ is again a point in the small region $\cal R$).  We have
therefore two results.  First, the detection probability does not
depend on the duration of $\cal R$ for such regions.  This is in fact
the key reason for which the duration of a measurement can be safely
neglected in the quantum theory of a particle: if this duration is
sufficiently short, the probability to detected the particle is
independent from such a duration.  Second, the probability is
proportional to the spatial extension of the interval.  Therefore we
can define a spatial probability density $\rho(x)$ by
\begin{eqnarray}
\rho(x) = \lim{}_{\Delta L\to 0}\ \  \frac{P_{\cal R}}{\Delta L},
\end{eqnarray}
and we obtain that the probability density in space is 
\begin{eqnarray}
\rho(x) =   |\psi(x)|^2. 
\end{eqnarray}

Cover a spatial region $\Omega$ at time $T$ by many contigous but non
overlapping small rectangular spacetime regions ${\cal R}_n$ of the
type we have been considering.  If we assemble the detectors of the
corresponding characteristic states $|{\cal R}_n\rangle$, we can ask
what is the probability that the particle will be detected by any of
the detectors.  Detection by different detectors are mutually
exclusive alternatives since their associated characteristic states
are orthogonal: $\int_{\cal R} d^2x \int_{\cal R'} d^2y W(x;y) \simeq
0$ when $\cal R$ and $\cal R'$ are simultaneous, non-overlapping, and
both have dimensions obeying $\Delta T \ll m\Delta X^2/\hbar$.  Thus
the probability is just the sum of the individual detection
probabilities.
\begin{eqnarray}
P = \sum_n P_{{\cal R}_n} 
	 \simeq & \sum_{n} (|\psi|^2 \Delta X)_n
	 \rightarrow & \int_\Omega |\psi|^2 dX,
\end{eqnarray}
where the last expression is exact in the limit of infinitesimal
$\Delta X$, and is accurate as long as $\psi$ is approximately
constant in each ${\cal R}_n$.  What we have found is of course
precisely the standard probability interpretation of the wavefunction. 
When applied to the non-relativistic quantum mechanics of a single
particle using the propagator specific to this system, the covariant
probability interpretation (\ref{eq:main}) yields the standard
probability interpretation of the wavefunction of this system.

It is important to notice that the fact that detection in disjoint
spatial regions at equal time are mutually exclusive alternatives does
not reflect a special role of time in the {\em formalism} but rather
is a feature of the propagator of the non-relativistic particle.

Finally, can we define the probability density in spacetime to find 
the particle around a spacetime point $x$? The answer is no, for the 
following reason. In order to be able to define such a probability, 
the following limit should exist 
\begin{eqnarray}
\tilde\rho(x) = \lim{}_{{\cal R}\to x}\ \ \frac{P_{\cal R}}{
V_{\cal R}},
\label{limit}
\end{eqnarray}
where $V_{\cal R}$ is the volume of the region.  Consider a region
$\cal R$ of sides $\epsilon L$ and $\epsilon T$.  A tedious integral
shows that for small $\epsilon$
\begin{eqnarray}
     C_{\cal R}^{-2} \simeq \sqrt{\hbar\over m} L^2 T^{3/2} 
     \epsilon^{7/2}.
\end{eqnarray}
Thus 
\begin{eqnarray}
P_{\cal R} = |\langle {\cal R}|\Psi\rangle|^2  =  C_{\cal R}^2 V_{\cal R}^2
 |\psi(x)|^2 = |\psi(x)|^2   (\epsilon T)^{-1/2},
\end{eqnarray}
so that the probability does not scale with the volume, and the limit 
(\ref{limit}) does not exist.  Therefore there is no probability 
density in spacetime. 

\subsection{Towards a covariant formulation of quantum theory}

The possibility of labeling quantum states by means of spacetime
functions $f({\X},{\T})$, or more generally functions of the
configuration space and time, opens the possibility of formulating
ordinary quantum theory in a form in which the time variable plays a
less peculiar role than in the conventional formulation.  To prepare
the ground for such a reformulation we begin here by reviewing the
structures we have introduced in the context of the non-relativistic
free particle in a more abstract mathematical manner, so that they can
then be generalized.

First, let $\cal H$ be the linear space of the physical solutions
$\psi({\X},{\T})$ of the Schr\"odinger equation.  $\cal H$ is
canonically isomorphic to the Hilbert space ${\cal H}_{0}$ of the
states $\Psi({\X})$ at fixed time $\T=0$: the identification map
$I:{\cal H}\rightarrow {\cal H}_{0}$ is given by the restriction
$\Psi({\X})=\psi({\X},0)$ and its inverse $I^{-1}:{\cal
H}_{0}\rightarrow {\cal H}$ is given by the Schr\"odinger evolution
\begin{equation}
\psi({\X},{\T}) = \int d{\X}'\ W({\X},{\T};{\X}',0)\ \Psi({\X}')
\label{evolution}
\end{equation}
The identification map induces the physical Hilbert product on $\cal
H$ by
\begin{equation}
(\psi,\psi')_{\cal H} \equiv (I\psi,I\psi')_{{\cal H}_{0}} = 
\int d{\X}\ \overline{\psi({\X},0)}\psi'({\X},0). 
\end{equation}

Next, consider a space $\cal E$ formed by spacetime ``test functions"
$f({\X},\T)$.  For concreteness, we take these functions to be smooth
and with compact support.  More general functions such as rapid
decrease functions, can be more convenient for some applications.  A
very important object is the linear map $P$ from the space of the test
functions ${\cal E}$ to the Hilbert space ${\cal H}$,  defined by
\begin{eqnarray} 
P\, : {\cal E} &\to& {\cal H} \\ 
 \ \ \ \ : f &\mapsto& |f> \nonumber \\ 
\psi_{f}({\X},{\T}) \hspace{5pt} \equiv  
\hspace{5pt} <{\X},{\T}|f>  &\equiv& \int d{\X}' d{\T}'\ 
  W({\X},{\T};{\X}',{\T}')\ 
  f({\X}',{\T}'). \nonumber 
\end{eqnarray} 
(See equation (\ref{mapP}).)  This map is highly degenerate: it
sends arbitrary functions in solutions of the Schr\"odinger equation. 
Its image is dense in ${\cal H}$.  The scalar product can be 
pulled back to $\cal E$, giving 
\begin{equation}
    \langle f|f'\rangle = \int d{\X}\ d{\T} \int
    d{\X}'\ d{\T}'\ \overline{f({\X},{\T})}\
    W({\X},{\T};{\X}',{\T}')\ f'({\X}',{\T}').
\label{bform1bis}
\end{equation}
(See equation (\ref{bform1}).)  And therefore the Hilbert space of the
theory $\cal H$ is nothing but the linear space $\cal E$ equipped by
the bilinear form (\ref{bform1bis}) (divised by the zero norm subspace
and completed in norm).  Therefore {\em the propagator $W(\x,\y)$
contains the full information needed to reconstruct the Hilbert space
of the theory from the linear space $\cal E$}.

Let $C$ be the Schr\"odinger operator defined on $\cal E$
\begin{equation} 
    C = \imath\hbar {\partial\over\partial \T}
    + {\hbar^{2}\over 2m}{\partial^{2}\over\partial {\X}^{2}},  
    \label{eq:Sch2} 
\end{equation}
Can $W(\x,\y)$ be recovered directly from the operator $C$ (without
passing via the $|\X,\T\rangle$ states as we did in section
\ref{free})?  The answer is positive, and there is a number of general
techniques to derive $W(\x,\y)$ directly from $C$.  Here we describe a
general technique denoted group averaging, essentially following 
Marolf's ideas \cite{Marolf}.  We refer the reader to
\cite{Marolf} and references therein for more details and a for more
complete mathematical treatement of the technique.  The operator $C$
defines on $\cal E$ the bilinear form
\begin{equation}
    (f',f)_{C} \equiv \int_{-\infty}^{\infty} d\tau \int d{\X} d{\T}\
    \overline{f'({\X},{\T})}\ \left[e^{i\tau C} f\right]\!({\X},{\T}).
   \label{bform}
\end{equation} 
This can be easily computed by Fourier transforming, obtaining
\begin{eqnarray}
    (f',f)_{C} &=& \int_{-\infty}^{\infty} d\tau \int d{\X}d{\T}
    d{\X}' d{\T}' dp dE\ e^{-ip{\X}'}\ e^{-iE{\T}'} \ \tilde f(-p,-E)
    \ \ e^{i\tau (E+p^{2}/2m)}\ e^{ip{\X}}\ e^{iE{\T}}\ \tilde f(p,E)
    \nonumber \\ &=& \int dp dE \ \tilde f(-p,-E)\ \delta(E+p^{2}/2m)\
    \tilde f(p,E) \nonumber \\
    &= & \int d{\X}'d{\T}' d{\X}d{\T}\ \overline {f'({\X}',{\T}')}\ 
    W({\X}',{\T}',{\X},{\T})\ f({\X},{\T}).
\end{eqnarray}
Therefore the propagator $W(x,y)$ is nothing by the kernel of the
bilinear form $(\ ,\ )_{C}$ defined in (\ref{bform}).  In turn, this
bilinear form is precisely the scalar product. Once more, we have:  
\begin{eqnarray}
    (Pf',Pf)_{\cal H} &=& \int d{\X}\ \overline{\psi_{f'}({\X},0)}\
    \psi_{f}({\X},0) \nonumber \\ &=& \int d{\X}
    d{\X}'d{\T}'d{\X}'{}'d{\T}'{}'\ 
    \nonumber \\ && \ \ \ \   \times \ \   
    \overline{W({\X},0;{\X}'{\T}')
    f'({\X}',{\T}')}\ W({\X},0;{\X}'{\T}')\  f'({\X}'{}',{\T}'{}')
    \nonumber \\ &=& \int d{\X}'d{\T}'d{\X}'{}'d{\T}'{}'\
    \overline{f'({\X}',{\T}')}\ W({\X}',{\T}';{\X}'{\T}')\
    f'({\X}'{}',{\T}'{}') \nonumber \\ &=& (f',f)_{C}.
\end{eqnarray}
Therefore $P$ maps isometrically the linear space $\cal E$ equipped
with the bilinear form $(\ ,\ )_{C}$ into the Hilbert space $\cal H$
of the theory.  As $P{\cal E}$ is dense in $\cal H$, it follows that
$\cal H$ is entirely determined by $\cal E$ and $(\ ,\ )_{C}$.

This is a remarkable conclusion, because one often finds in the
literature the statement that in order to define the scalar product on
the space of the solutions of the Schr\"odinger equation one has first
to identify $\T$ as the time variable.  Contrary to this statement, we
see here that, at least for this simple case, the Hilbert space of the
theory {\em including its scalar product structure\/} is entirely
determined over a space of functions on spacetime by the Schr\"odinger
operator $C$, without having to single out the variable $\T$ as
``special".

\section{General covariant quantum theory}\label{general}

Let us now leave the simple case of a free particle, and consider the
general situation.  Consider a classical dynamical system.  The
kinematics of the system is defined by an (extended) configuration
space ${\cal M}$.  We call $x$ the points in ${\cal M}$, and we assume
that a measure $dx$ is fixed.  The dynamics of the system is defined
by a (single for simplicity) constraint $C=0$.  Here $C$ is a function
on $\Gamma$, where $\Gamma=T^{*}{{\cal M}}$ is the (extended) phase
space, namely the cotangent space of ${\cal M}$.  The couple $({\cal
M},\ C)$ completely defines the system.  We call here this formulation
of classical dynamics ``covariant".  Other denominations in the
literature are presymplectic, parametrized, extended \ldots

It is well known that a conventional hamiltonian system can be cast in
this form.  A conventional hamiltonian system is formulated in terms
of a configuration space ${\cal M}_{ph}$ and a hamiltonian $H$ which
is a function on the phase space $\Gamma_{Ph}=T^{*}{{\cal M}}_{ph}$. 
The hamiltonian generates evolution in an external time variable $\T$. 
To reformulate this system in covariant form, one promotes $\T$ to a
configuration space variable: the extended configuration space
includes the conventional configuration space plus the time $T$.  That
is, one poses ${{\cal M}}={{\cal M}}_{ph}\times I\!\!R$, where the
coordinate of $I\!\!R$ is identified with ${\T}$.  Also, one poses
$C=p_{T}+H$, where $p_{T}$ is the variable conjugate to ${\T}$ (which
physically turns out to be minus the energy).

A (well known) crucial observation is that most interesting physical
systems, and in particular all gravitational systems, such as full
general relativity with or without matter, cosmological models \ldots\
are not given in terms of a hamiltonian: they are given directly in
the covariant formulation.  Therefore not only the covariant
formulation of mechanics appears to be more general than the
hamiltonian one, but such a wider generality is required for the
theories that better describe our world.

One can try to ``deparametrize" these theories by picking one of the
configuration space space variables and identifying it as the time
variable.  It is sometimes claimed that such a deparametrization is
necessary in order to understand the quantum properties of the these
systems.  But such a deparametrization adds an element of
arbitrariness which is certainly not part of the classical dynamics. 
Since the classical dynamics of these systems does not select any
preferred independent ``time" variable $\T$, we think that their
quantum mechanics should not select a preferred time variable either. 
To understand their quantum dynamics, we must therefore have a
formulation of quantum theory in which time plays no special role. 
This is the motivation for the definition of covariant quantum theory
that we give in this section.

We want thus to quantize the system $({\cal M},\ C)$.  We begin with a
space $\cal E$ of test functions $f(\x)$ over ${\cal M}$.  (We use now
latin letters $\x,\y,\ldots$ for points in the extended configuration
space.)  The quantum dynamics is then determined by a propagator
$W(x,y)$ on ${\cal M}\times{\cal M}$.  

Here we are more interested in the interpretation of the theory once
the propagator is given, than in the actual construction of the
propagator.  Let us nevertheless say something on the derivation of
$W(x,y)$ itself.  There is a number of ways of constructing this
object starting from the classical theory.  For instance, $W(x,y)$ may
be defined as a sum over classical histories \cite{Hartle}.  In the
case of nonperturbative quantum gravity, $W(x,y)$ may be defined by
means of an auxiliary quantum field over a group \cite{spinfoam}.  In
a canonical quantization, assume that an operator $C=C(\x,-i\hbar
{\partial\over\partial \x})$ whose classical limit is the constraint
$C$ is given (that is, assume a given operator ordering has been
chosen).  One can then follow closely Marolf's construction
\cite{Marolf} mentioned in the previous section.  That is, define a
bilinear form on $\cal E$
\begin{equation}
    (f',f)_{C} \equiv \int_{-\infty}^{\infty} d\tau 
    \int d\x\ \overline{f'(\x)}\ \left[e^{i\tau C} f\right]\!(\x). 
   \label{bform2}
\end{equation} 
as in equation (\ref{bform}).  See \cite{Marolf}.  The kernel of this
bilinear form defines $W(x,y)$:
\begin{equation}
   (f,f')_{C} = \int d\x d\x'\ \overline{f(\x)} \ W(\x,\x')\ f(\x'). 
    \label{eq:W}
\end{equation}

Once $W(x,y)$ is determined, the rest of the formalism and the
interpretation of the quantum theory follow.  The Hilbert space of the
theory is defined as the space $\cal H$ obtained by equipping $\cal E$
with the bilinear form (\ref{eq:W}), quotienting by the kernel of the
bilinear form and completing in norm.  The physical states can
therefore be labelled by the functions on ${{\cal M}}$ as $|f\rangle$. 
The (highly degenerate) map
\begin{eqnarray} 
P\, : {\cal E} &\to& {\cal H} \\ 
 \ \ \ \ : f &\mapsto& |f>  
\end{eqnarray} 
is often improperly called the ``projector".   

Each state $|f\rangle$ determines a solution of the quantum constraint
equation via
\begin{equation} 
 \psi_{f}({\X},{\T}) \equiv \int d{\X}' d{\T}'\ 
  W({\X},{\T};{\X}',{\T}')\ 
  f({\X}',{\T}'). 
\end{equation} 
The scalar product on the space of these solutions, is well defined by
$\langle f|f'\rangle =(f,f')_{C}$, namely by (\ref{eq:W}).  But in
general it may not be easilly written directly in terms of the
solutions $\psi_{f}({\X},{\T})$ themselves.

We must then give the theory an interpretation.  Associate a
(normalized) state $|{\cal R}\rangle$ to each finite region ${\cal R}$
in ${{\cal M}}$.  The state $|{\cal R}\rangle$ is defined as
\begin{equation}
    |{\cal R}\rangle = C_{\cal R}\ |f_{\cal R}\rangle, 
    \label{eq:defR}
\end{equation}
where $f_{\cal R}$ is the characteristic function of the region ${\cal
R}$ and $C_{\cal R}=|\langle f_{\cal R}|f_{\cal R}\rangle|^{-1/2}\ $ is
the normalization factor.  We define the interpretation of the theory
by \emph{postulating} that if ${\cal R}$ is sufficiently small (namely
in the limit in which ${\cal R}$ is still finite, but smaller than any
other physically relevant dimension involved in the problem), the
probability amplitude to detect a system (in the state $|\Psi\rangle$)
in the region ${\cal R}$ is given by
\begin{equation} 
    P_{\cal R} =|\langle {\cal R}|\Psi\rangle|^2
    \label{eq:main2}
\end{equation}
In turn, $|{\cal R}\rangle$ represents the state of the system after a
measurement that has detected the system in the region ${\cal R}$. 

In particular, the quantity 
\begin{equation}
    A_{{\cal R},{\cal R}'} =\langle {\cal R}|{\cal R}'\rangle
    \label{eq:main2bis}
\end{equation}
is the probability amplitude to detect the system in the (small)
region $\cal R$ of the extended configuration space, if the system was
previously detected in the (small) region ${\cal R}'$.  This amplitude
can be written explicitely in terms of the propagator as
\begin{equation}
    A_{{\cal R},{\cal R}'} = 
     C_{\cal R}
      C_{{\cal R}'} \int_{\cal R}dx\int_{{\cal R}'} dy \ 
    W(x;y) \ =\  
      \frac{\int_{\cal R}dx\int_{{\cal R}'} dy \ 
    W(x;y)}{\sqrt{\int_{\cal R}dx\int_{{\cal R}} dy \ 
    W(x;y)}\ \ \sqrt{\int_{{\cal R}'}dx\int_{{\cal R}'} dy \ 
    W(x;y)}}.
    \label{eq:main3}
\end{equation}
This completes the definition of general covariant quantum theory.

Whether or not the limit ${\cal R}\to\x$ can be taken, as well as any
peculiar property of the probabilities in this limit, depends on the
dynamics.  Notice that the limit of the amplitude $ A_{{\cal R},{\cal
R}'}$ as ${\cal R}$ shrinks to $\x$ and ${\cal R}'$ to $\x'$ is
proportional to $W(\x,\x')$.  Therefore $W(\x,\x')$ is proportional to
the amplitude for the system to be at $\x$ if it was at $\x'$. 
However, the proportionality factors $ C_{\cal R}C_{{\cal R}'}$ are
dimensional and may diverge in the limit.  Furthermore the divergence
may depend on the way the limit is taken.  For instance, it may depend
on the \emph{shape} of the region approaching the point $x$.

\section{Applications}\label{applications}

We now sketch the application of the general formulation discussed
above to a number of physical systems, of increasing complexity.

\subsection{Non relativistic particle}

Does the general theory defined in Section \ref{general} agrees with
the conventional quantum theory of a non-relativistic particle studied
in Section \ref{free}?  The interpretation of the states $|{\cal
R}\rangle$ is precisely the same as the one we have derived for the
free particle, and therefore is okay.  What about the interpretation
of $\langle {\cal R}'|{\cal R}\rangle$, for sufficiently small
regions, as a transition amplitude?  Let us distinguish three cases,
according to whether ${\cal R}'$ is entirely in the past of ${\cal
R}$, entirely in the future, or neither.

First, we exclude the last case from our considerations, on the ground
that the postulated interpretation demands ${\cal R}$ and ${\cal R}'$
to be smaller than any other relevant dimension in the problem.  In
order for ${\cal R}$ and ${\cal R}'$ not to be time ordered, their
relative time localization must be of the same order than their size.

Second, consider the case in which ${\cal R}'$ is in the future of
${\cal R}$.  In this case, the amplitude for detecting in ${\cal R}'$
a particle prepared in ${\cal R}$ is indeed proportional to $\langle
{\cal R}'|{\cal R}\rangle$, where the proportionality factor depends
on the efficiency of the detector.  Therefore the interpretation
suggested agrees with the prediction of conventional quantum
mechanics.

Finally, consider the case in which ${\cal R}'$ is in the past of ${\cal R}$. 
Strictly speaking, this case refers to a situation which has no
meaning in conventional quantum mechanics: it refers to a situation in
which the measurement is made at an earlier time than the preparation. 
Therefore the general theory given in Section \ref{general} gives more
predictions than the ones usually considered in conventional quantum
theory.  The additional predictions can be more accurately denoted
``retrodictions", since they are statements about a time which is in
the past with respect to the time at we assume to have information
about the state.  Jim Hartle has long argued that such retrodictions
can be added to standard predictions of quantum theory, and in fact,
that they have to be added, if we want to make sense of any statement
about the past deriving from our knowledge of the present.  Either we
give up the possibility of making {\em any\/} statement about the
past, or we take retrodictions as these statements.  We refer to
Hartle's paper \cite{Hartle} for a detailed discussion.

Whatever position we take about retrodiction, the usual predictions of
the quantum theory of a non relativistic particle are recovered from
the general formalism of Section \ref{general}.  The extension from
the free case to the case with a potential is immediate.

\subsection{Time of arrival}

If the reader is not interested in this problem, this section can be
skipped without prejudice for understanding what follows.  A simple
application of the above considerations is to the problem of the time
of arrival in quantum theory.  This problem has generated considerable
discussions \cite{toa}.  As far as we understand, there is no
agreement on its solution.  The problem is the following.  Suppose
that at $\T=0$ a nonrelativistic particle is in the state $\Psi(\X)$. 
A particle detector is placed at the origin.  At what time $\T$ will
the particle detector detect the particle?  More precisely, how can we
compute the probability distribution in time $\rho(\T)$ that the
detector detects the particle at time $\T$?  Surprsingly, there is no
agreement on the solution of this simple problem of non relativistic
quantum mechanichs, in spite of the fact that the problem can
presumably be experimentally investigated.  Different authors have
computed different distributions $\rho(\T)$ !

The considerations in the previous section suggest the following
answer to the problem.  First, no detector can be concentrated at the
origin.  Second, the time resolution of the particle detector can only
be finite.  Therefore, the only probabilities that we can realistic
hope to measure are the probabilities that the particle be measured in
a small but finite spatial interval $I$ around the origin, in a small
but finite time interval of size $\epsilon$ around the time $\T$.  Let us
imagine that the particle detector is placed around the origin.  It
detects the particle in the finite region $I$, and it has a discrete
set of pointer variables $i_{n}$.  The $n$-th pointer variable $i_{n}$
indicates that the particle has been detected by the apparatus between
the times $\T_{n}=n\epsilon$ and $\T_{n+1}=(n+1)\epsilon$.  In other words, in
spacetime language, we consider a collection of detectors.  The $n$-th
detector is a spacetime detector of finite resolution $I\times \epsilon$
around the spacetime point $\X=0$, $\T=n\epsilon$.  We denote this
spacetime region as $R_{n}$.

We can then associate a two-state model detector of the kind described
in the Section \ref{realistic} to each region $R_{n}$, and the
computation of the probabilities for their final configuration at some
later time $\T_{fin}$ is a standard exercize in quantum theory.

The important interpretational point is the following.  Does one of
the detectors detects whether at the time $\T$ the particle is at the
space point $\X$, or does this detector detects whether, at the
origin, the passage time of the particle is $\T$?  In other words, is
this a position measurement at a given time or a time measurement at a
given position?  Is it a projector on an eigenspace of the position
operator $\X(\T)$ or of a time of arrival operator $\T(\X)$?  The
answer is that a realistic detector is neither of the two.  It is an
approximation to both quantities.  {\em If we take into account that
realistic measurements cannot have infinite time resolution, the
distinction between measurement of position at a certain time and
measurement of time of arrival at a certain position, disappears}.

\subsection{Relativistic particle, I}\label{I}
 
The quantum theory of a single relativistic particle is not a
realistic theory, since it neglects the physical phenomena of particle
creation which are described by quantum field theory.  Nevertheless
it is interesting to ask whether there exist a logically consistent 
quantum theory, or several, whose classical limit is the
dynamics of a single relativistic particle, and which respects the
Lorentz invariance of the classical theory.  We discuss two such
quantizations.  In the first we consider only positive frequency
solutions of the Klein Gordon equation, in the second, we consider a
theory for both frequencies.

We start with the following covariant formulation of the classical
theory.  We take ${\cal M}$ to be Minkowski space, and the constraint
$C$ to be given by the two conditions 
\begin{eqnarray}
    p^{2} & = & m^{2},
    \label{eq:p2} \\   
    E & > & 0. 
    \label{eq:p0}
\end{eqnarray} 
where $p=(P,E)$ and $p^{2}=-P^{2}+E^{2}$.  (We use here $\hbar=c=1$.) 
Upon quantization the constraint (\ref{eq:p2}) becomes the
Klein Gordon equation and positive energy condition (\ref{eq:p0}) 
becomes the restriction to positive frequencies. Together these constraints
restrict the wavefunction to be the Fourier transform of a function
supported on the upper mass hyperboliod in momentum space.
The corresponding propagator is
\begin{eqnarray}
  W(x,y)=\int \frac{d^{2}p}{2\pi}\ \delta(p^{2}-m^{2})\ \theta(E)\ e^{-ip(\x-\y)}
  = \int \frac{dP}{2\pi}{1\over 2E(P)}\ e^{iP(\x^{1}-\y^{1})-iE(P)(\x^{0}-\y^{0})}, 
\end{eqnarray}
where $E(P)=+\sqrt{P^{2}+m^{2}}$, and $\x= (\x^{1}, \x^{0})=
(\X,\T) $.  It is easiest to represent the Hilbert space in momentum
space.  Let
\begin{equation}
  \tilde f(p)= \int d^{2}x \ f(x)\ e^{-i(p\cdot x)}. 
\end{equation}
We have then 
\begin{equation}
   (f,g)_{C}=\int \frac{d^{2}p}{2\pi}\ \delta(p^{2}-m^{2})\ \theta(E) \bar{\tilde f(p)}\
   \tilde g(p). 
\end{equation} 
Therefore the state space can be represented by  ${\cal H}=L_{2}[R^{2},
\delta(p^{2}-m^{2})\ \theta(E)\ d^{2}p]/\cal I$, the quotient of the space 
of functions on $R^2$ whose restrictions to the upper mass hyperboloid are 
square integrable with the Lorentz invariant measure, by the zero norm 
subspace $\cal I$ - the subspace of functions which vanish almost everywhere 
on the upper mass hyperboloid.  
Clearly the states can be expressed in terms of functions on the upper
mass shell only, which (once an inertial frame is chosen) can be written
as functions of the spatial momentum $P$ only: The Hilbert space is then
${\cal H}=L_{2}[R,{dP\over2E(P)}]$, with the wavefunction corresponding to 
$\tilde f$ being $\Psi(P)=\tilde f(P,E(P))$ and the inner product given by
\begin{equation}
   (\Psi,\Theta)_{C}=\int \frac{dP}{2\pi}\: \frac{1}{2E(P)}\: \bar{\Psi(P)}\
   \Theta(P). 
\end{equation} 

Historically two types of (generalized) states have been associated to
spacetime points $\x=(\X,\T)$ in relativisitic quantum mechanics.  
First \cite{phil} there is the Philips state $\Phi_{\x}$, which we also 
denote $|\x\rangle$, which is the spacetime smeared state defined by the
spacetime delta function $f(y) =\delta^2 (y-x)$:
\begin{equation}
    \Phi_{\x}(P) = \langle P|\x\rangle = e^{-i(P\x^{1}-E(P)\x^{0})}. 
\end{equation}
Second \cite{NW}, there is the Newton-Wigner state $\Psi_{x,K}$, which we 
also denote $|x,K\rangle$:
\begin{equation}\label{NWstate}
    \Psi_{x,K}(P)= \langle P|x,K\rangle = \sqrt{2E(P)}\
    e^{-i(P\X-E(P)\T)}.
\end{equation} 
The Philips state depends only the spacetime point $\x$, while the
Newton-Wigner state depends on $x$, as well as on the choice of an
inertial frame $K$, the frame to which the energy $E(P)$ and all other
space and time components in the expression (\ref{NWstate}) are refered.  
The solutions of the Klein-Gordon equation corresponding to these states
are, respectively,
\begin{equation}
    \phi_{x}(y)= W(\y,\x) 
\end{equation}
and 
\begin{equation}
    \psi_{x,K}(y)= \int dT'{}'\ W(y,\X,\T'{}')\ 
    v(\T-\T'{}'), 
\end{equation} 
where (reinstating dimensionful constants and normalizing the Newton-Wigner state
with a factor $(mc^2)^{-1/2}$),  
\begin{equation}
         v(\T) 
	 = \int \frac{dE}{2\pi\hbar}\ \sqrt{\frac{2|E|}{mc^2}}\ e^{iE\T/\hbar} 
	 = -\frac{1}{2\sqrt{\pi}\tau_c} \left(\frac{\tau_{c}}{|T|}\right)^{-3/2}
\end{equation} 
with $\tau_{c}=\frac{\hbar}{mc^2}$ the ``Compton period" of the particle. 

It is easy to see that two Newton-Wigner states, corresponding to two 
distinct points but the same inertial frame and the same time, are orthogonal.  
Indeed
\begin{equation}
   \langle (\X,\T), K|(\X',\T), K\rangle = \delta(\X,\X'). 
\end{equation}
Furthermore, as we vary $\X$ at fixed $\T$ these states span the
Hilbert space, forming a basis.  There exists, therefore, a selfadjoint 
position operator $X_K(T)$ diagonal in this basis -
\begin{equation}
    X_K(T)\ |(\X,\T), K\rangle = X |(\X,\T), K\rangle,
\end{equation} 
- which is called the Newton-Wigner position operator.  At
$\T=0$ its form in the representation $\Psi(P)$ adapted to the frame $K$ is
\begin{equation}
     X_K = i\sqrt{E(P)}\ {d\over d P}\ {1\over 
     \sqrt{E(P)}}. 
\end{equation}

The Newton-Wigner states are defined with respect to a particular inertial 
frame.  The state $|(\X,\T), K\rangle$ is not a spatial position 
eigenstate with respect to the position operator of another frame.
Moreover, $|x, K\rangle $ and $|x', K\rangle$ are not orthogonal unless 
$x \neq x'$ are simultaneous in the frame $K$. It is not sufficient
that they be spacelike separated.
The Phillips states do not depend on a choice of reference frame and
their inner products $\langle x|y\rangle = W(x,y)$ show that they are 
not quite orthogonal for {\em} any spacelike separated $x$, $y$. 

Let us now see how these results appear from the point of view of the
general theory of Section \ref{general}.  Let us consider a small
region ${\cal R}$ centered in the points $\x$, and study the limit of
the state $|{\cal R}\rangle$ as $R\to\x$.  We can take for instance a
rectangular region and scale its spacial and temporal sides as
$L\to\epsilon L$ and $T\to\epsilon T$, adjusting the normalization of 
the state
appropriately.  If the normalized characteristic function approximates
a 2 dimensional delta function, then the state we obtain approximates
the Phillips state of the point, namely $|\x\rangle$.

The states $|x,K\rangle$ can also be obtained as limits of characteristic
states, using  regions of a more complicated shape.  For instance, 
the Newton-Wigner state centered on the spacetime origin can be obtained
using the region defined by $|\X| <\epsilon \left(
\frac{\tau_{c}}{T} \right)^{3/2}$.  For small $\epsilon$, we obtain a
state that approximates $|(0,0),K\rangle$.  Of course, the asymptotic
points of this thin diamond shaped region in Minkowski space pick out
the axes of a specific Lorentz frame.  Hence the frame dependence 
of the Newton-Wigner states.

Can detectors be built that detect these states?  In each Lorentz
frame, the theory is just like the non relativistic particle theory
with hamiltonian $H=+\sqrt{P^2+m^2}$.  We can therefore use the same
detector described for the non relativistic particle, with
appropriately shaped interaction regions.  In all this, of course, one
should keep in mind that the theory is not realistic, since reality is
described by quantum field theory.  For detectors corresponding to the
Newton-Wigner operator, see also \cite{Marolf}.
Therefore the two kinds of states correspond simply to the "point-like"
limits of two distinct kinds of measurements. Notice that the detectors 
corresponding to the Newton-Wigner states associated with a single equal 
time surface collectively measure the Newton-Wigner position, with the
eigenvalue obtained being the position of the (only) detector that finds
the particle.

Both states propagate faster than light, in the sense that there is a
finite probability that the particle be detected at two spacelike
separated points.  There is nothing logically inconsistent in this: it
is simply a prediction from this quantum theory.  In the classical
limit, the trajectories stay inside the light cone.

Notice that there are two distinct ways of characterizing a state
$|\psi\rangle$ by means of a function of the position at fixed time
$\T=0$ in a frame $K$.  First, we can take $\Phi(\X)=\langle x|\psi\rangle$ 
with $\x=(\X,0)$.  Second, we can take $\Psi(\X)=\langle 
(\X,0),K|\psi\rangle$.  The first is the the value of the solution 
$\psi(\X,\T)$ of the Klein-Gordon equation corresponding to the state 
at $\T=0$, while the second is the amplitude of finding the particle in 
$\x=(\X,0)$ by means of a Newton-Wigner position measurement in frame $K$.  
The two quantities are distinct.  Both characterize the state uniquely.  
They are related by
\begin{eqnarray}
    \Phi(\X) & = & \int \frac{dP}{2\pi} {1\over 2E(P)}\ \Psi(P)\ e^{iP\X}; 
    \nonumber  \\
    \Psi(\X) & = & \int \frac{dP}{2\pi} {1\over \sqrt{2 E(P)}}\ \Psi(P)\ e^{iP\X}. 
\end{eqnarray}

\subsection{Relativistic particle, II}\label{II}

A different quantum theory for a single relativistic particle is
obtained by dropping the positive frequency condition (\ref{eq:p0}). 
The Hilbert space $\cal H$ is then formed by functions with support on
both hyperboloids.  The propagators is
\begin{eqnarray}
  W(\x,\y)&=&\int \frac{d^{2}p}{2\pi}\ \ \delta(p^{2}-m^{2})\ e^{ip(\x-\y)}
  \nonumber \\
&=& \int \frac{dP}{2\pi}{1\over 2E(P)}\ \left(e^{iP(\x^{1}-\y^{1})-iE(P)(\x^{0}-\y^{0})}
+ e^{-iP(\x^{1}-\y^{1})+iE(P)(\x^{0}-\y^{0})} \right).
\nonumber
\end{eqnarray} 
Again, this is proportional to the probability amplitude of finding
the particle in $\x$ if it was in $\y$.  The key difference between
this theory and the one in the previous subsection is the fact that in
this theory the spatial localization of the particle in a given frame
is not sufficient to determine the state.  Indeed, a function $\psi(p)$ 
on {\em both} hyperboloids is not determined by a single function of 
the spatial momentum, but rather by two functions of momentum:
\begin{equation}
    \Psi_{\pm}(P)\equiv\psi(P,\pm E(P)).
    \label{eq:pm}
\end{equation}
We can still define the Newton-Wigner position operator at $\T=0$ in
a given reference frame
\begin{equation}
     X \Psi_{\pm}(P) = -i\hbar\ \sqrt{E(P)}\ {d\over d P}\
     {1\over \sqrt{E(P)}}\ \Psi_{\pm}(P),
\end{equation}
but this operator no longer constitutes a complete set of commuting
observables.  Its (generalized) eigenspaces are doubly degenerate.  They
include a positive frequency and a negative frequency component. 
Accordingly, the position of the particle at $\T=0$ does not determine
the state uniquely.  

In a sense, the theory describes a particle that can exist in two
states: either as a particle or as an antiparticle.  The dynamics does
not mix the the two, but a measurement does.  Different kinds of
measurements can select different mixtures of positive and negative
frequency states.  

A measurement that checks whether the particle is in a small region
${\cal R}$ defines the state $|{\cal R}\rangle$, represented by the
(non-normalized) solution of the Klein-Gordon equation
\begin{equation}
    \psi_{\cal R}(\x)= \int_{\cal R} W(\x,\y)\ d\y.     
\end{equation}
If a particle is in such a state, the probability amplitude of finding
it in a small region ${\cal R}'$ is given by (\ref{eq:main3}).

\subsection{Quantum particle in a curved spacetime}

Let $g_{\mu\nu}(x)$ be a globally hyperbolic spacetime metric.  Can we
define a quantum theory of a single particle moving in the spacetime
defined by this metric?  The quantization of Section \ref{I} cannot be
generalized to a curved spacetime because in general there is no split
of the space of the solutions of the curved space Klein-Gordon
equation into positive and negative frequencies.  However, the
quantization of Section \ref{II} can.  The classical particle is
characterized by the constraint
\begin{equation}
    p^{2}\equiv g^{\mu\nu}(x)\ p_{\mu}p_{\nu} =  m^{2}. 
\end{equation} 
where $p=(P,E)=(p_{1},p_{0})$.  The quantum constraint becomes the
curved spacetime Klein-Gordon equation
\begin{eqnarray}
   C\  \psi(\x) = (g^{\mu\nu}(x)D_{\mu}D_{\nu} - m^{2})\ \psi(\x)=0. 
\end{eqnarray} 
where $D_{\mu}$ is the covariant derivative of $g$.  The state space
if formed by solutions of this equation, and the propagator $W(x,y)$
is defined by
\begin{eqnarray}
  (f,f')_{C} &=& \int d\tau \int d^{2}\x\ \overline{f(\x)}\ \left[e^{i\tau 
  C}f\right]\!(\x)
  \nonumber \\
  &=& \int d^{2}\x d^{2}\y\ \overline{f(\x)}\ W(\x,\y)\ f(\y), 
\end{eqnarray}
assuming that the integral in the first line converges. 

As before, a measurement that checks whether the particle is in a
small spavcetime region ${\cal R}$ defines the state $|{\cal
R}\rangle$, represented by the solution of the Klein-Gordon equation
\begin{equation}
    \psi_{\cal R}(\x)= \int_{\cal R} W(\x,\y)\ d\y.     
\end{equation}
If a particle is in such a state, the probability amplitude of finding
it in a small region ${\cal R}'$ is given as before by (\ref{eq:main3}). 

\subsection{Quantum cosmology}

We do not consider here the problem of making sense of the
quantum theory of a single universe, in which the frequency
interpretation of probabilities is questionable, and in which the
notion of external observer, required by the Copenhagen
interpretation, is of difficult use.  Instead, we assume that the
Wheeler DeWitt equation considered does not describe all degrees of
freedom of the universe, but only a subset of these (say the
gravitational ones, or just some cosmological variables), so that we
can still assume, for the sake of the interpretation, that other
degrees of freedom in the universe are treated classically, and can be
used to define a classical Copenhagen external observer.  And also
that the dynamics that we are studying is such that in some
appropriate sense measurements could be repeated on same states, and
thus the frequency interpretation of probability could be used. 
Whether or not these assumptions are physically viable is a problem we
do not address here. We focus only on the issue of time.  

We assume we are given a Wheeler DeWitt equation, of the form
$C\psi=0$ as a differential equation for a wave function $\psi(\x)$,
where $x\in {{\cal M}}$ represents a set of physical variables.  The
space ${{\cal M}}$ can be infinite dimensional, for instance the space
of the three geometries, or finite dimensional.  For instance, in a
simple homogeneous isotropic cosmological model with a scalar field,
we have $\x=(a,\phi)$, where $a$ is the radius of the universe and
$\phi$ is the spatially constant value of the scalar field.  We focus
here on the finite dimensional case, since we are interested in the
conceptual issue of time only, leaving the generalization to an
infinite dimensional ${\cal M}$ to further developments.  We also
assume that we can fix a measure $d\x$ on ${{\cal M}}$ giving an
auxiliary Hilbert space $H_{aux}=L_{2}[{{\cal M}},d\x]$ in which $C$
is self adjoint, and a space $\Phi$ formed by smooth compact support
functions $f(x)$ on $M$.  The question is whether we can give a
consistent probabilistic interpretation to the solutions of the
Wheeler DeWitt equation without selecting in ${{\cal M}}$ a time
variable, or a preferred time direction.

Our strategy should be clear at this point.  We define the bilinear
form $(\ , \ )_{C}$ on $\Phi$ in terms of Equation (\ref{bform2}).  We
mod out by the zero norm states and complete in norm, obtaining an
Hilbert space $\cal H$.  This can be identified as a space of
solutions of the Wheeler DeWitt equations.  Any function $f$ in $\Phi$
defines a state $|f\rangle$ of the system.  In particular, for any
small but finite region ${\cal R}$ in ${\cal M}$ we consider the state
$|{\cal R}\rangle$ and we give it the physical interpretation of a state
which has been found in the region ${\cal R}$ of the extended configuration
space.  The probability to find the system in the region ${\cal R}'$ is then
given by Eq.\ (\ref{eq:main2bis}).

\subsection{Quantum gravity}

In reference \cite{ac}, a strategy for computing transition amplitudes
$W(s,s')$ between (generalized) three geometries $s,s'$ is given.  The
space on which the generalized three geometries $s$ live is a
collection of finite dimensional components and carries a natural
measure $ds$.  We can therefore interpret the theory along the lines
discussed here.  The theory defined probabilities of finding a certain
three geometry (with matter), within a small error, after a certain
three geometry (with matter), within a small error, has been detected.

\section{Some conceptual issues}\label{issues}

The introduction of the spacetime-smeared observables raises certain 
general issues, which we discuss here. 

\subsection{Probability of what?}

Consider the following objection 
\begin{itemize}
    \item The probability of finding the particle in the space
    interval $I$ at $\T=0$ is meaningful because the alternative is
    well defined: it is the probability that the particle be
    elsewhere, namely in $I\!\!R-I$, $I\!\!R$ being the $\T=0$ real
    line.  But the probability of the measurement outcome of finding
    the particle in an arbitrary spacetime region ${\cal R}$ has no meaning,
    because the set of alternatives is not defined. 
\end{itemize}

Consider a measurement of position at fixed time $\T$, on an assigned
initial state $\Psi$, and assume here that we can perform the
measurement with infinite time resolution.  Let us say that the
probability that particle is in the interval $I$ is $P$.  This
probability can be measured (to a given accuracy) by repeated
measurements, as relative frequency of outcomes.  If we talk about
frequency, we have to specify the set of alternatives out of which the
outcome is considered.  Otherwise, the notions of probability or
frequency do not make sense.

In a measurement of position, the alternative is often taken to be
that the particle is elsewhere (not in $I$) at $\T=0$.  Thus, $P$ is
interpreted as the probability that the particle is $I$, as opposite
of being in $I\!\!R-I$.  To make sense of this definition of
alternatives, one should assume that we have an infinity of detectors
spread all along the real line $\T=0$, all the way to infinity: some
detectors on Andromeda, some on Orion, and others further away.  There
is nothing wrong in idealizations, but is this an useful one?  Is this
idealization needed to make sense of the measurement of the
localization of a particle?  In a concrete experiment, what we do is
simply to turn on the detector, and see whether it has detected the
particle.  Why should this be related to the behavior of another
ideal particle detector on Andromeda?  

It is more reasonable to assume that the alternatives we consider are
whether this particular detector has detected the particle {\em or
not.} We can consider a set of two alternatives only: one is that the
detector detects the particle, the other is that the detector does not
detect the particle.  Once the initial state of the particle is
determined and the detector is specified, these two alternatives are
well defined and form a complete set of alternatives.  The frequency
interpretation of measurement positions of quantum particles in a real
experiment, for instance, is clearly this one, and not whether the
particle is in $I$ as opposite at being somewhere else.

In the case of a non relativistic particle, the {\em theory\/} tells
us that if the particle is detected in $I$, then it cannot be detected
in $I\!\!R-I$.  This, however is a consequence of the dynamics of the
theory, not an a priori requirement needed to make sense of the
measurement.

Consider now a small spacetime region ${\cal R}$ and a detector there. 
We can consistently define the probability that the particle is in
${\cal R}$ as the probability that the detector detects the particle. 
This is not the probability that the particle is ${\cal R}$ as
opposite of being somewhere else.  It is the probability that the
particle be detected, as opposite of not being detected.

\subsection{What is an observable? Partial observables and complete 
observables}

Consider the following objection
\begin{itemize}
  \item In quantum mechanics, the position of the particle $\X$ is an
  observable, while the time $\T$ is an external parameter.  One
  should not confuse the two, which are very distinct.  In particular,
  if we say that we measure whether the particle is in ${\cal R}$, we
  are assuming that we can measure quantum mechanically both position
  and time, and this is a mistake, because time is not an observable.
\end{itemize}

This point is discussed in detail in reference \cite{partialobs}. 
Here we give a short account of the response, referring the reader to
\cite{partialobs} for more details.  There is a certain ambiguity in
the notion of observable.  This ambiguity is reflected for instance in
the difference between the quantum theory observables in the
Shr\"odinger picture (the ``position" operator $\X$) and in the
Heisenberg picture (the ``position at time $\T$" operator $\X(\T)$.) 
To disentangle this ambiguity, let us start from the classical
mechanics of a single particle.  At every time $\T$, we can measure
the position $\X$ where the particle is.  Let us use the expression
``complete observable" to indicate the position of the particle at a
given time.  Thus, a complete observable is for instance the position
at $\T=0$, and a distinct complete observable is the position at
$\T=3$ seconds.  We use the expression ``partial observable" to
generically indicate the ``position" or the ``time".  More precisely,
we operationally define a partial observable as any measurement
procedure that produces a number (checking where is the particle,
looking at the clock\ldots).  We define a complete observable as a
measurements procedure that gives a number that can be predicted from
the knowledge of the state of motion of the system (or, in quantum
theory, whose probability distribution can be predicted.  A typical
complete observable is formed by the conjunction of two (or more)
partial observables.

In non-covariant theories, partial observables fall in two distinct
groups: independent and dependent ones.  Independent partial
observables characterize the spacetime position where the measurement
happens.  Thus, time is the independent partial observable in the
mechanics of a nonrelativistic particle, while the position of the
particle is the dependent one.  In Maxwell theory, time and position
are two independent partial observables, the electric and magnetic
field are the dependent observables, and a complete observable is
given by the value of the field at a given time and a given position.

The novelty introduced with general relativity, and the peculiarity of
all covariant theories is precisely the fact that the a priori
distinction between dependent and independent partial observables is
lost.  Complete observables are still given by conjunctions of partial
observables, but dependent and independent partial observables are not
distinguished.  Thus for instance in the cosmology of a isotropic
universe with a constant scalar field, $a$ and $\phi$ are both partial
observables, but they are on an equal footing.

Back to the objection, the truly observable quantity is the relation
between $\X$ and $\T$.  That is, a state of motion determines a unique
relation between $\X$ and $\T$.  Because of the specific form of the
dynamics of the system, we can then treat the two quantity
dissymmetrically.  We can treat $\T$ as an external independent
parameter and $\X$ as a dynamical variable.  In a general case,
however, this may be impossible.  A general state of motion will
determine a relation between a set of variables in an extended
configuration space ${\cal M}$.  Thus, the distinction a priori
between $\X$ and $\T$ mentioned in the objection is viable, but not
necessary, in a non covariant theory.  Is not anymore viable in a
covariant context.

\subsection{Repeated measurements}

Up to now, we have considered the situation in which quantum theory is
employed to predict the probability for the outcome of a measurement
on a state prepared in another measurement.  Standard quantum theory,
however, has a wider application: it can be applied to repeated
measurements.  That is, quantum theory addresses the following
problem.  Assume we know that the system is in a state $|\Psi\rangle$;
let a first measurement be performed and let us assume that we know
the outcome $A$ of the measurement.  What is the state after the
measurement?  Equivalently, how can we compute the probability $P$ of
obtaining $A$ {\em and\ } obtaining $B$ in a another measurement
(of a different observable)?

The standard answer is the following.  If $\Pi_{A}$ is the projector on
the eigenspace corresponding to the outcome $A$, and $\Pi_{B}$ is the
projector on the eigenspace corresponding to the outcome $B$, then the
probability of obtaining $A$ and $B$ is $P_{BA} = |\Pi_{B}\Pi_{A} \Psi
\rangle|^{2}$ if the $A$ measurement is the first in time.  It is
$P_{AB}=|\Pi_{A}\Pi_{B}\Psi\rangle|^{2}$ if the $B$ measurement is the
first in time.  In general, the two projectors $\Pi_{A}$ and $\Pi_{B}$ do
not commute, and therefore $P_{AB}$ is not equal to $P_{BA}$. 
Therefore the probability of a set of outcomes is determined by the
{\em time ordering\ } of the measurements.  

If we try to analyze this situation in our covariant framework we find
that the framework it is not complete, since time ordering is not
defined in a covariant theory.  This fact raises a difficulty.  To see
this, consider three regions of the extended configuration space:
${\cal R}$, ${\cal R}'$ and ${\cal R}'{}'$.  Let the state be $|{\cal
R}\rangle$.  What is the probability that the system is detected in
${\cal R}'$ {\em and\ } in ${\cal R}'{}'$?

It is tempting to say that the measurement of the system in ${\cal
R}'$ prepares a state $|{\cal R}'><{\cal R}'|{\cal R}> \equiv
\Pi_{{\cal R}'}|{\cal R}>$ on which the ${\cal R}"$ measurement acts
and so the probability is $||\Pi_{{\cal R}"} \Pi_{{\cal R}'} {\cal
R}>|^2$.  But which projector, $\Pi_{{\cal R}'}$ or $\Pi_{{\cal R}"}$,
should be applied first?

In the case of a non-relativistic particle being detected in two very
small spacetime regions ${\cal R}'$ and ${\cal R}"$ with ${\cal R}"$
to the future of ${\cal R}'$ and both to the future of ${\cal R}$, the
formula $P = ||\Pi_{{\cal R}"}\Pi_{{\cal R}'}{\cal R}>|^2$ does indeed
reproduce the correct standard result.  On the other hand, the
quantity $ P=|\Pi_{{\cal R}'}\Pi_{{\cal R}"} R\rangle|^{2}$ does not seem
to have any clear physical meaning in the theory \cite{Hartle2}.  But
if it is so, how can we generalize this result to an arbitrary
configuration space ${\cal M}$, in which no time ordering is defined? 
The projectors act on the wavefunctions in all of spacetime, including
the past, so it is not obvious that the projector corresponding to the
later measurement should be put on the left.  Indeed it may not be
possible to define which region is later.  For instance in a special
relativistic context even disjoint regions may be partly in the future
and partly in the past of each other.

There are two, possibly related approaches.  In the first approach
\cite{Marolf3} the series of measurements is treated as a single
measurement by including part of the measurement aparatus in the
system of observation, with the reading of the results of the sequence
of measurements by the observer constituting the final, single act of
measurement.

For instance, in the example above assume that in the region ${\cal
R}'$ the particle interacts with a two state system $S_{1}$ and in the
region ${\cal R}'{'}$ with a two state system $S_{2}$, as in section
\ref{realistic}.  The extended configuration space ${\cal M}$ is the
physical spacetime $M$ times $\{0,1\}$ times $\{0,1\}$.  Assume the
initial state is characterized by the region ($R, 0, 0)$ of ${\cal
M}$.  Let us ask what is the probability that the system is found in
$(R_{\T}, 1, 1)$, where $R_{\T}$ is the $\T=constant$ line in
spacetime and $\T$ is later than ${\cal R}'$ and ${\cal R}'{}'$.  It
is then easy to see that the resulting probability amplitude is
proportional to $ A=|T(\Pi_{{\cal R}'}\Pi_{{\cal R}'{}'}) R\rangle|^{2}$,
where $T$ indicates time ordering.  That is, time ordering is not
produced by an additional postulate of the quantum theory, but simply
by the dynamics itself.  The quantum theory predicts only outcomes of
individual measurements.  A sequence of measurements can be
reformulated as a single measurement, by including into the system the
dynamics of the measurement apparatus.

The second answer is more speculative.  In order to use quantum
theory, we ideally separate the world into two components.  The first
component is the system studied, which we denote $S$.  The second
component is the ``observer", namely the rest of the world, which we
denote $O$.  We think that this separation is intrinsic in the quantum
description of the world: quantum theory is always a theory of the
interactions of a system ($S$) over another system
($O$).\footnote{However, there are well developed attempts to make
sense of the quantum theory of ``closed" systems, namely systems that
do not interact with an external observer.  See in particular
\cite{Hartle}.} On the other hand, the division is arbitrary: indeed,
the most remarkable feature of quantum theory is that the descriptions
obtained by breaking the world into $S/O$ components in different
manners are all consistent with each other \cite{qm}.  Now, the $S$
system may be a covariant system in which time ordering is not
defined.  Nevertheless, a time ordering may be introduced by $O$.  In
other words, the time ordering that selects the relevant probabilities
may be the one of the observer, not the one of the system.  For
instance, imagine that $S$ is formed by the $a$ and $\phi$ degrees of
freedom in a cosmological situation.  We can take then as $O$ a set of
variables describing physics on, say, the Earth, for which a specific
time ordering is somehow physically determined.  A sequence of
measurements in the ${\cal M}$ extended configuration space with
coordinates $a$ and $\phi$ is then time ordered by the order under
which the system $O$ comes in relation with these regions.  

\section{Conclusion}\label{concl}

We have explored the possibility of defining quantum theory in a
covariant form.  That is, in a form that allows the independent time
variable to be treated on an equal footing with the dynamical
variables in the extended configuration space.  We expect that such a
form of quantum theory is required for understanding the quantum
behavior of the covariant systems such as the relativistic
gravitational systems.

We have found that much of the peculiar role that the time variable
assumes in the conventional formulation of quantum theory is not
intrinsic in the quantum behavior of the physical systems, but rather 
it  depends on an idealization of the measurements: the
unrealistic assumption that physical measurements could be performed
instantaneously.  This idealization simplifies the formalism of quantum
theory; however, it hides the beautiful symmetry among all variables
of the extended configuration space.  This symmetry is present in
classical mechanics, where it is made manifest by formulations such as
the Hamilton-Jacobi theory, or the covariant (presymplectic,
parametrized) hamiltonian formulation.  The thesis of this paper is
that this symmetry is not broken by the physical quantum
phenomenology, but only by the unrealistic idealization of
instantaneous measurements.  Giving up this idealization reveals the
same symmetry in the quantum world, opening the way to a formulation
of quantum theory sufficiently general to deal with covariant
theories.

The technical ingredient to be added to the quantum formalism is the
notion of spacetime-smeared quantum state.  This is a state generated
by a measurements that is not instantaneous.  In particular,
localization measurements can be naturally described in terms of
states associated to spacetime regions, or, more in general, regions
in the extended configuration space.  The key element of the theory,
from this point of view, is the propagator $W(\x,\y)$.  This quantity
is a two point function on configuration space, it is closely related
(but not identical!)  to the ``probability amplitude for the system to
be detected in $\x$ if it was detected in $\y$".  Furthermore, it
defines the Hilbert space of the theory, since it is the kernel of the
scalar product between spacetime-smeared states, and it defined the
``projection" from the space of test functions over configuration
space to the Hilbert space itself.  In general, $W(\x,\y)$ is defined
by the Wheeled DeWitt operator.  In the classical limit, it is easy to
see that $W(\x,\y)$ reduces to the exponential of the classical
action, or, more precisely, to a general solution of the
Hamilton-Jacobi equation of the system.

The formulation we have suggested makes sense for non relativistic
systems, at the (interesting) price of adding the ``retrodictions" to
the predictions of the quantum theory.  It makes sense in the context
of (unrealistic) relativistic theories of a single particle, where it
helps clarifying the distinction between the different kind of states
associated to spacetime points (such as the Newton-Wigner and the
Phillips states).  It allows us to make sense of the (unrealistic)
quantum theory of a single particle in a curved spacetime.  It allows
us to give a logically consistent interpretation to the (realistic?) 
quantum cosmological models, as far as the ``problem of time" of these
models is concerned.  We expect also that this general formulation can
be taken as a reference scheme in quantum gravity.

The remaining conceptual difficulty regards the possibility of
associating probabilities to sequences of measurements.  We see two
possible solutions to this difficulty.  The first by reducing any such
sequence to a single measurement or, equivalently, to sets of
commuting measurements, by including the apparatus in the theory.  The
second by introducing the notion of time ordering of the observer.

\centerline{------------}

We thank Rafael Porto and Don Marolf for valuable exchanges on the
subject, Bruno Iochum and Raphael Zentner for numerical help with an
integral.  This work was partially supported by NSF Grants
PHY-9900791.

{\em Note added}: In the recent paper \cite{DC}, the interpretation
postulate presented here is shown to be correct in the context of the
dynamics of a relativistic particle, provided that a Lorentz invariant
description of the measurement is utilized.

\end{document}